# NeoGuard: a public, online learning platform for neonatal seizures

Amir Hossein Ansari [1,2] *, Perumpillichira Joseph Cherian [3,4], Alexander Caicedo [1,2], Anneleen Dereymaeker [5], Katrien Jansen [5,6], Leen De Wispelaere [7], Charlotte Dielman [8], Jan Vervisch [5], Paul Govaert [7,8], Maarten De Vos [9], Gunnar Naulaers [5], Sabine Van Huffel [1,2]

[1] Department of Electrical Engineering (ESAT), KU Leuven, Leuven, Belgium

[2] imec, Leuven, Belgium

[3] Section of Clinical Neurophysiology, Department of Neurology, Erasmus MC, University Medical Center Rotterdam, The Netherlands

[4] Division of Neurology, Department of Medicine, McMaster University, Hamilton, Canada

[5] Department of Development and Regeneration, University Hospitals Leuven, Neonatal Intensive Care Unit, KU Leuven, Leuven, Belgium.

[6] Department of Development and Regeneration, University Hospitals Leuven, Child Neurology, KU Leuven, Leuven, Belgium

[7] Section of Neonatology, Department of Pediatrics, Sophia Children's Hospital, Erasmus MC, University Medical Center Rotterdam, The Netherlands

[8] ZNA Koningin Paola Kinderziekenhuis, Antwerp, Belgium

[9] Institute of Biomedical Engineering, Department of Engineering, University of Oxford, Oxford, UK

∗ Corresponding author

E-mail: amirhossein.ansari@kuleuven.be

# Abstract

Seizures occur in the neonatal period more frequently than other periods of life and usually denote the presence of serious brain dysfunction. The gold standard for detecting seizures is based on visual inspection of multichannel continuous electroencephalogram (cEEG) complemented by video analysis, performed by an expert clinical neurophysiologist. Previous studies have reported varying degree of agreement between expert EEG readers, with kappa coefficients ranging from 0.4 to 0.85, calling into question the validity of visual scoring. This variability in visual scoring of neonatal seizures may be due to factors such as reader expertise and the nature of expressed patterns. One of the possible reasons for low inter-rater agreement is the absence of any benchmark for the EEG readers to be able to compare their opinions. One way to develop this is to use a shared multi-center neonatal seizure database and use the inputs from multiple experts. This will also improve the teaching of trainees, and help to avoid potential bias from a single expert's opinion. In this paper, we introduce and explain the NeoGuard public learning platform that can be used by trainees, tutors, and expert EEG readers who are interested to test their knowledge and learn from neonatal EEG-polygraphic segments scored by several expert EEG readers. For this platform, 1919 clinically relevant segments, totaling 280h, recorded from 71 term neonates in two centers, including a wide variety of seizures and artifacts were used. These segments were scored by 4 EEG readers from three different centers. Users of this platform can score an arbitrary number of segments and then test their scoring with the experts' opinions. The kappa and joint probability of agreement, is then shown as inter-rater agreement metrics between the user and each of the experts. The platform is publicly available at the NeoGuard website (www.neoguard.net).

# Introduction

Seizures commonly denote underlying brain injury in neonates and are generally associated with high mortality and morbidity. The prevalence of clinical seizures in newborn babies is estimated to be 2 to 3 per 1000 live births (0.2-0.3%), and is even higher (9-11%) in very low-birthweight or premature neonates.[1–3] The actual prevalence is likely to be underestimated as neonatal seizures are often subtle or subclinical, and clinical observation alone is known to be unreliable in their diagnosis.[3] Hypoxic-ischemic encephalopathy (HIE) is one of the most common and serious causes of neonatal seizures (30-53%). Other etiologies include metabolic disturbances, intracranial bleeding, infections, cerebral infarction as well as malformations. Neonatal seizures are brief lasting (few seconds to minutes) events of excessive and abnormal neuronal electrical activity and usually are regional or focal. Generalized tonic-clonic seizures are rarely seen in the first month of life.[3,4] Furthermore, the majority of neonatal seizures are clinically subtle or silent, so-called 'electrical only (electrographic) seizures', mandating the use of EEG monitoring for their accurate diagnosis and estimation of seizure burden. It is known that clinical observation alone results in marked underestimation of seizure burden, even by trained observers.[3,5–8]

One of the best-known and widely used EEG-based monitors is the cerebral function monitor (CFM$^{TM}$), which uses a compressed, rectified version of the EEG, called amplitude integrated EEG (aEEG) recorded using 2 to 4 EEG channels. The CFM generally displays about 5 hours of recording in one page and is used in many neonatal intensive care units (NICUs) because of its ease of use. However, it has been reported that aEEG often misses brief-duration (<30s), low-amplitude, or focal/regional seizures. Furthermore, some types of artifacts can easily be misinterpreted as seizures in aEEG, especially by non-experts.[9,10] Particularly in neonates with severe encephalopathy, treatment with high dose of anti-epileptic drugs (AEDs) may suppress EEG signals and result in brief-duration, low-amplitude seizures, which may be poorly detected by CFM.

Multichannel Continuous EEG monitoring (cEEG) is the ideal method for detecting seizures. Generally, visual inspection of multichannel cEEG complemented by video is considered to be the gold standard for diagnosing neonatal seizures, although their interpretation needs special knowledge and training. Moreover, cEEG is expensive and labor-intensive as compared to CFM.[9,11–13] One of the main challenges confronting the interpretation of EEG is the poor agreement of expert EEG readers for some types of seizures,[14,15] especially for 'dubious seizures'. Dubious seizures are low-amplitude, arrhythmic, paroxysmal EEG events lasting shorter or longer than 10s, with no regular evolution of amplitude, frequency, and morphology. Such seizures commonly occur in neonates with severe encephalopathy or after treatment by high dose AEDs.[15] Seizure patterns with a higher disagreement among raters would also result in higher variability in treatment decisions. Furthermore, studies reporting automated neonatal seizure detection methods have highlighted the challenge of distinguishing brief-duration seizures from brief-lasting artifacts.[16–18] Lower inter-rater agreement between expert readers is also one of the hurdles for developing reliable automated seizure detection methods, when one realizes that the so-called "ground truth" is ambiguous.

It is likely that expert EEG readers working in one center have generally higher agreement compared to experts from other centers. The lack of a shared benchmark for neonatal seizure detection and scoring is a major issue. EEG trainers mostly use their local datasets for teaching trainees. This could contribute to the variability in pattern recognition and interpretation, especially when the characteristics of the patient populations differ. Creating a public platform for self-testing and comparing these tests with the opinion of different experts is essential for trainees and tutors. Such a platform can also be used by experts to share information, better define core characteristics of neonatal seizures, develop consensus and ultimately lead to improved standards of EEG interpretation, and impact quality of medical care and research.

Recently, an online website, called NeoGuard Information System, has been developed by our research group and consists of three modules: 1) an EEG database, for collecting and storing EEG data from the participating teaching hospitals, 2) a scoring system, for presenting and scoring the EEG segments by our expert readers, and finally 3) the NeoGuard learning platform (NGLP), for sharing our experts' scores with trainees and other experts. The third module, NGLP, is described in this paper. The main goal of the NGLP is to provide a platform teaching users how to annotate subclinical seizures from neonatal EEG-polygraphic recordings. This platform is open to researchers/clinicians with different levels of experience. The platform enables users to learn and compare their scores with the ones provided by our experts for EEG segments including definite seizures, dubious seizures, or seizure-like artifacts. In this paper, the NGLP is introduced, its usage is described, and its added value, as well as limitations, are discussed.

## Materials and methods

### Database

The EEG segments used in this platform were obtained from 71 neonates recorded at two centers: 1) the NICU of Erasmus MC, University Medical Center Rotterdam, the Netherlands (EMCR) and 2) the NICU of the University Hospitals Leuven, Belgium (UZL). All recorded data were filtered between 1 and 20 Hz before being fed into the automated seizure detection algorithm or rescored by the secondary raters. All recordings were anonymized in the corresponding hospitals. This work was conducted with the approval from the Ethical Review Boards in both hospitals.

### 1) EMCR

EEG-polygraphic data of 48 term neonates were recorded in this center between 2003 and 2012. All neonates were term (with gestational age ≥ 36), admitted to the NICU with presumed postasphyxial HIE, and underwent cEEG and MRI. The inclusion criteria for asphyxia were either a five minute Apgar score below 6 or arterial pH of the umbilical cord blood less than 7.1 and clinical encephalopathy according to Sarnat score. Treatment with anti-epileptic drugs was initiated by the protocol explained by Cherian et al.[19] if either electrographic or electroclinical seizures were detected. The neonates with heart or brain malformation were excluded and no other preselection was made. For 35 neonates, a full 10-20 International System using 17 electrodes, Fig 1 (A), was used. In recordings from 10 neonates, the electrodes $F_{3-4}$ and $P_{3-4}$ were not available and a restricted system using 13 electrodes, Fig 1 (B), was used. For the remaining 3 neonates, electrodes $F_{7-8}$ and $T_{5-6}$, in addition to $F_{3-4}$ and $P_{3-4}$, were not recorded, hence a restricted system using 9 electrodes, Fig 1 (C), was used.[19] The polygraphic signals included electrocardiogram (ECG), electro-oculogram (EOG), chin or limb surface electromyogram (EMG), and abdominal respiratory movement signal (Resp). All data were recorded by a Nervus[TM] monitor (Taugagreining hf, Reykjavik, Iceland) or a Brain RT EEG system (OSG BVBA Rumst, Belgium). The sampling frequency was 256 Hz and the primary filtering was between 0.3 and 70 Hz. For this database, 8h each from 17 neonates, 4h each from 18 neonates, and 2h each from 13 neonates were selected by an experienced clinical neurophysiologist because of their high occurrence of seizures (in total 234h).

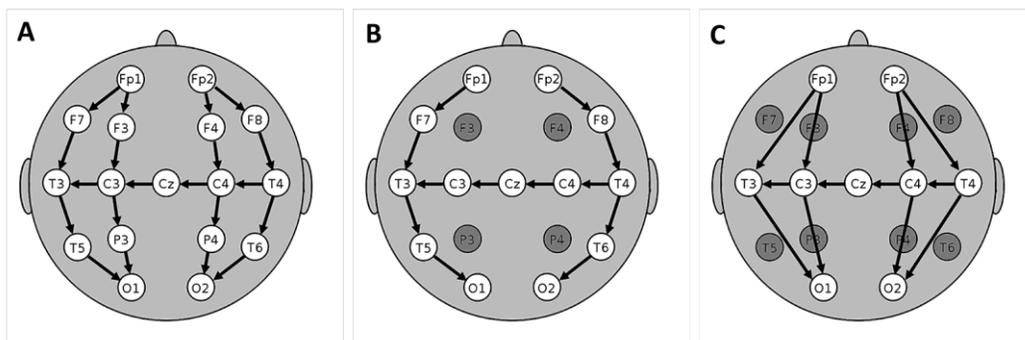

*Fig 1. Neonatal EEG montages. A: full 10-20 system of electrode placement using 17 electrodes, B) a restricted 10-20 system using 13 electrodes, C) a restricted 10-20 system using 9 electrodes.*

## 2) UZL

From this center, EEG data from 23 term neonates, who were admitted to the NICU and had seizures in the neonatal period, recorded between 2013 and 2015 were selected for this database. Among these, 6 neonates had HIE, 5 had metabolic problems, 5 had stroke, 2 had genetic malformations, and the remaining 5 had other etiologies. The neonates with heart malformations were excluded and no further preselection was performed. The EEG-polygraphic signals were recorded by a Brain RT EEG system (OSG BVBA Rumst, Belgium), using 9 EEG electrodes, $Fp_{1-2}$, $T_{4-5}$, $C_{4-5}$, Cz, $O_{1-2}$ placed according to the restricted 10-20 International System. Twelve bipolar channels were displayed,[19] Fig 1 (C). The polygraphic signals were the same as for the EMCR data, mentioned above. The data were recorded at 250 Hz initially and then resampled to 256 Hz. Two-hour EEG recordings from each neonate in which at least one seizure has been observed were selected and scored by an experienced clinical neurophysiologist (in total 46h).

## Rescoring strategy

The seizure segments were annotated in EMCR and UZL by two expert clinical neurophysiologists, one from each center. In total, 7% of recordings (25h out of 353h) were labeled as seizure. In order to construct the dataset of the training platform, some segments should be selected from the recordings and they should be re-scored by independent experts. If the segments are selected randomly from the recordings, it will lead to a very unbalance dataset (7% seizure vs. 93% nonseizure). Instead, in order to have more clinically relevant segments, all annotated seizures have been selected, as well as the nonseizure annotations which were misclassified by an automated algorithm. Thus, the selected nonseizure segments usually include artifacts, dubious seizures that were not annotated as seizure, or seizure-like patterns according to the automatic detector. To this end, a previously developed heuristic neonatal seizure detector mimicking a human observer, developed in,[20] was applied on the data. Then, the automatically detected segments were compared with the labeled seizures and, consequently, they were classified into two groups: falsely or truly detected segments. Next, the former group was merged and shuffled with the seizure group, forming the 'event pool'. Thus, the event pool includes all segments scored as seizure by the human experts, as well as all false alarms of a seizure detector (in total 1919 events). Finally, 3 independent expert EEG readers rescored all events and labeled them as 'definite seizure', 'dubious seizure', and 'definite artifact'. As a consequence, for each event, 4 scores are available, 1 provided by the primary and 3 provided by the secondary expert EEG readers. These labels are explained in the next section. Fig 2 schematically shows the mentioned scenario.

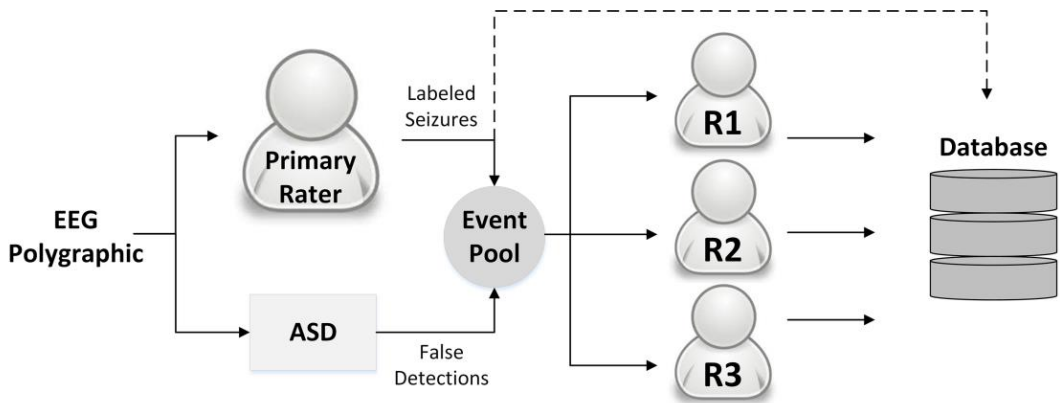

Fig 2. Schematic diagram of the multi-scoring scenario. R1-3 denote the 3 secondary raters and ASD is the automated seizure detector.

## Method

As mentioned in the previous section, the goal of the NGLP is providing a platform for trainees and expert EEG readers to compare their knowledge and experience with experts of our group. To this end, a website was developed and is made publicly available. Anyone can open the website, choose the number of events that he/she wants to score, define some of the EEG review settings, and start to score. Each of these steps is explained in the following sub-sections.

## Scoring labels

Each event can be scored as one of the following labels:

1) *Definite Nonseizure/Artifact*: our experts used this label when no pattern suspicious for seizure is present or when the EEG segment only expresses artifacts, such as those originating from the ECG, EOG, EMG, respiration, etc. Fig 3 shows a definite artifact which was scored as nonseizure by the primary clinical neurophysiologist and all secondary raters as well.

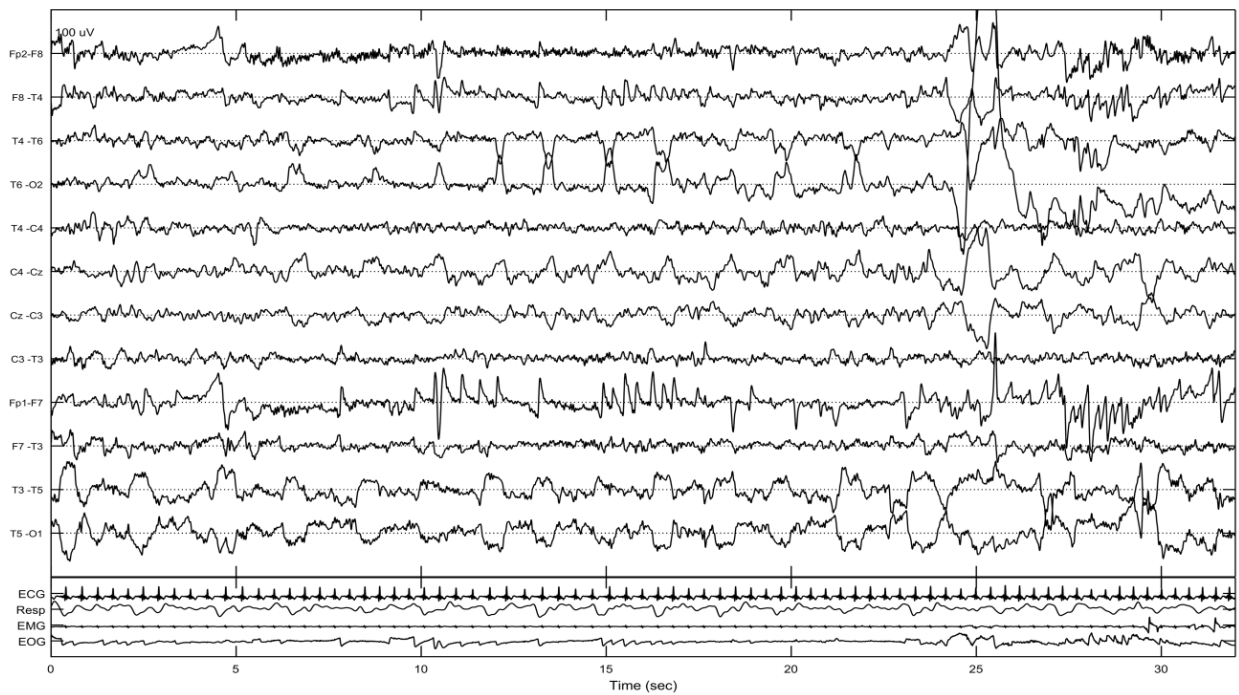

*Fig 3. A definite artifact that was scored as 'nonseizure' by the primary neurophysiologist and all secondary raters.*

2) *Definite seizure*: our experts used this label when there is a clear pattern of rhythmic oscillations, repetitive sharp waves, or mixture of both on the EEG channels which lasted more than 10 seconds. This pattern should have clear evolution of frequency, amplitude, or morphology and it should look different from the ongoing background activity. Furthermore, the rater believed that clinicians will initiate or become highly encouraged to initiate treatment with anti-epileptic drugs (AED) because of this event. Fig 4 displays a definite seizure lasting for 19s which was scored as seizure by the primary and all secondary raters.

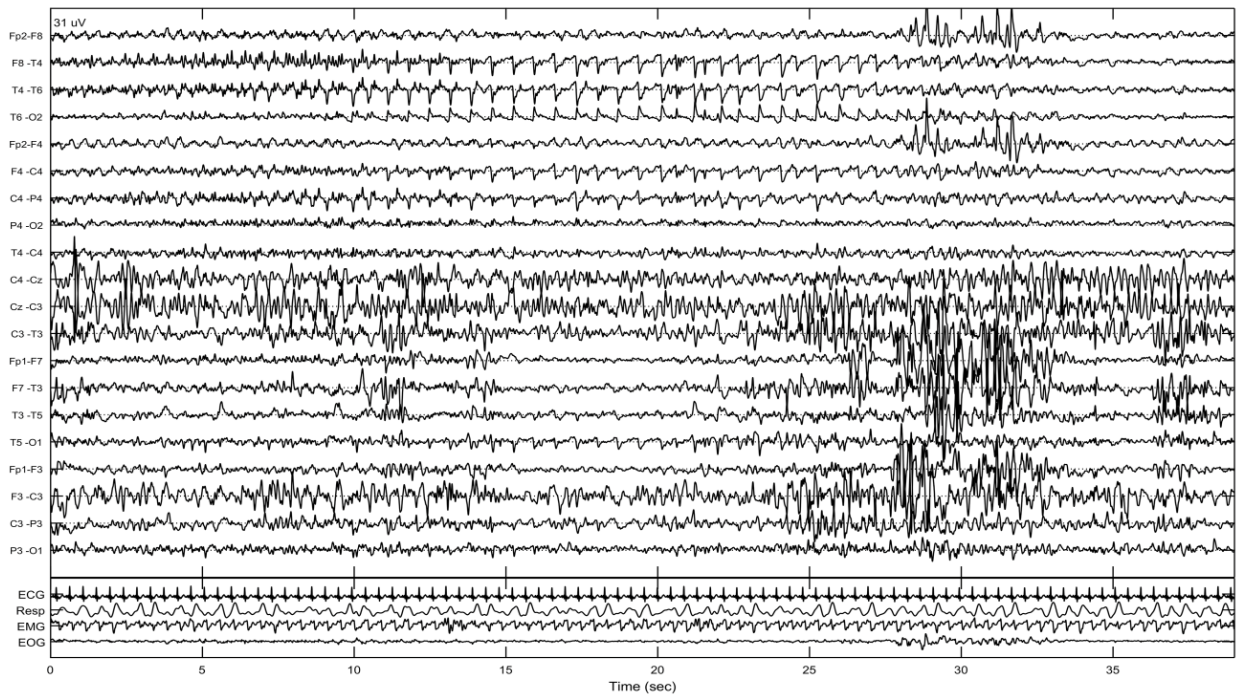

*Fig 4. A definite seizure, starting from sec 10 and lasting for 19s (see Methods for definition) over the right temporal region. This event was scored as 'seizure' by the primary neurophysiologist and all secondary raters.*

3) *Dubious seizure*: this label was used by our experts when there are dubious repetitive patterns in the EEG signal without sufficient characteristics of a definite seizure. This doubt usually occurs for very low-amplitude patterns with no clear evolution of amplitude, frequency, and morphology. They generally

(10)

appeared in neonates with serious brain dysfunction and usually was associated with poor clinical outcome. For this type of events, the clinicians will become convinced and initiate the AED treatment only if the pattern repeats itself and leads to a high seizure burden. Therefore, in the raters' view, a single dubious pattern should be considered neither as a seizure nor as a nonseizure/artifact. Fig 5 shows a dubious seizure detected by the primary neurophysiologist occurring at Cz electrode for 15 s with arrhythmic oscillations and no clear evolution of amplitude, frequency, or morphology. This event was scored by the three secondary raters as dubious, dubious, and definite seizure.

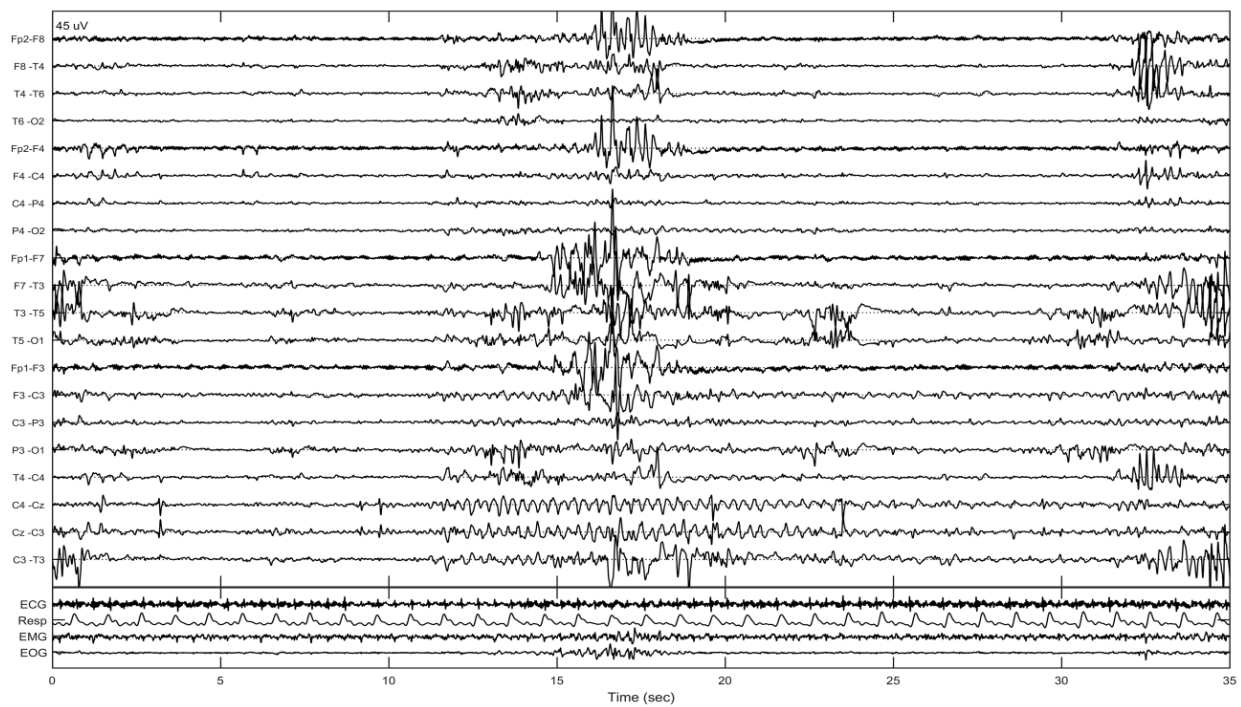

*Fig 5. A dubious seizure starting from sec 10 and lasting for 15s, with arrhythmic oscillations, without clear evolution in morphology, at Cz electrode, with an amplitude of ~25µv. This event was scored as 'dubious seizure' by the primary neurophysiologist. It was then rescored as (Dsz, Dsz, Sz) by the secondary raters.*

(11)

## Level of agreement

The events are partitioned into 3 classes based on the agreement of our expert raters. To this end, the most frequent scores of each event are counted and its percentage, *p%*, defines the agreement of each event. Based on *p*, the 3 classes are defined as follows: poorly agreed (*p ≤ 25%*), moderately agreed (*25% ≤ p ≤ 75%*), and highly agreed (*p ≥ 75%*). A user wanting to start scoring the events can choose the event class, or choose the 'Random' option, which does not take the agreement into account.

## Measuring raters' agreement

In order to score the user, the agreements between the user and our expert raters are measured by two inter-class correlation coefficients, which have commonly been used as inter-rater agreement metrics: I) joint probability of agreement, and II) Cohens' kappa. Both metrics are calculated between a couple of raters. The joint probability of agreement is computed as the number of events over which both raters agreed, divided by the total number of events. This metric does not take into account the fact that two raters may sometimes have similar opinions only by chance. The kappa is a corrected version of the joint probability of agreement and is calculated by

$$k = \frac{p_0 - p_e}{1 - p_e} \quad (1)$$

where $p_0$ is the joint probability of agreement and $p_e$ is the hypothetical expected probability of chance agreement measured by the following equation.

$$p_e = \sum_{j=1}^{N} p_j^1 p_j^2 \quad (2)$$

where $N$ is the total number of labels (=3 in this paper) and $p_j^r$ is the marginal probability of the $r^{\text{th}}$ rater for the $j^{\text{th}}$ class.[21,22] For example, assume that a rater scores 70%, 20%, and 10% of the events as 'Seizure',

(12)

'Dubious', and 'Nonseizure' respectively, whereas these values equal 60%, 10%, and 30% by the 2[nd] rater. In this case, $p_e$ equals $(70\% \times 60\%) + (20\% \times 10\%) + (10\% \times 30\%) = 47\%$. It shows that for 47% of these imaginary events, the raters may agree only by chance. As it is clear in (1), this chance probability is removed from the joint probability in the kappa statistic. Both these metrics are calculated between the user and each of our expert raters and reported at the end of the scoring process.

## Storing the users' Info

All scorings of expert or non-expert users are saved in the database of the NGLP. A user wanting to start the scoring process needs to agree that all entered inputs, including the user information (see Fig 6), as well as the scorings, will be saved in our database and may be used for future research.

## Start page

In order to start the scoring, the user should enter some initial information and settings. First, a CAPTCHA (Completely Automated Public Turing test to tell Computers and Humans Apart) security code should be entered and validated (Fig 6). CAPTCHA code is a worldwide Turing test to prevent automatic codes or malware from connecting to the website to make problems or steal the data. Second, the level of experience should be selected from 'Non-expert EEG Reader', 'Neurologist (Expert EEG Reader)', 'Neurophysiologist (Expert EEG Reader)', 'Pediatrician (Expert EEG Reader)', or 'Other (Expert EEG Reader)'. This information can help us to further refine the analysis of the current users' scorings. Third, the number of events that the user wants to score should be chosen (between 1 and 1000) and the level of agreement of upcoming events should be selected as explained. Users who would like to have us contact them may choose to enter their email or name. Finally, the user must allow our team to use the inputs for future analysis. Fig 6 shows a screenshot of the start page.

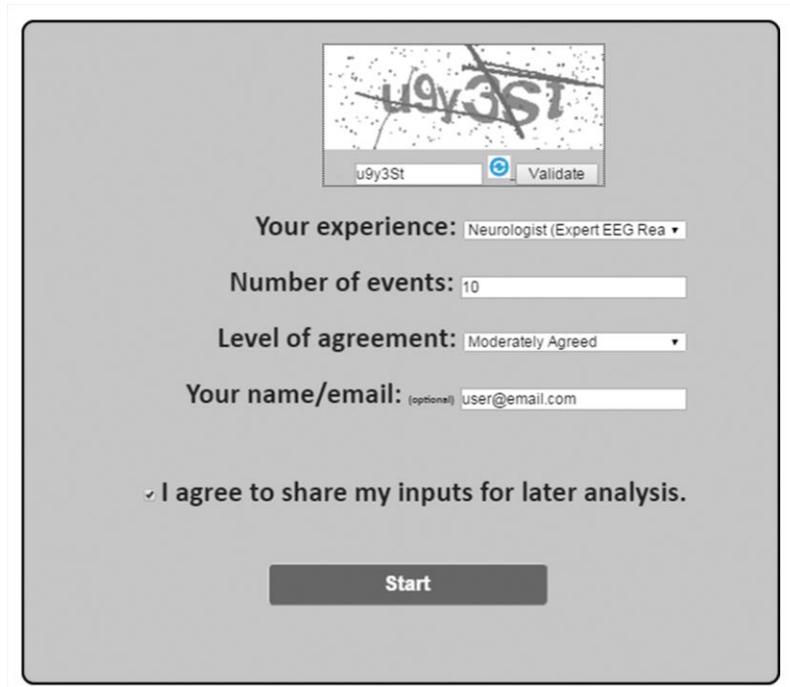

Fig 6. A screenshot of the NGLP start page

## Scoring page

The user scores each event based on 5 images as follows: image 1 displays the EEG-polygraphic signals of the whole event in seconds. Images 2, 3 and 4, show the details of signals in three 20sec-pages from 0-20sec, 20-40sec, and 40-60sec respectively. If the signal lasted more than 60sec, the detail images only show the beginning 60sec. On the other hand, if the signal lasted less than 20sec, images 3 and 4 will not appear (if the signal lasted less than 40sec, image 4 will not be available). Finally, image 5 gives a wider view of the event, starting 100sec before the start of the event and ending 100sec after the end of event. It can help the user to place the event in its context and help check whether it is an ictal event or part of a longer artifact. On top of the page, there are three buttons corresponding to the three labels: 'Artifact-

(14)

Nonseizure', 'Definite seizure', and 'Dubious seizure'. Furthermore, the user can choose the sensitivity of the EEG signals between '20uV', '50uV', '100uV', and '150uV', except for some specific events for which very small or very high sensitivity is needed. Some shortcuts are implemented for jumping between the images, scoring the events, and changing the sensitivity to speed up the scoring process. These are shown on this page. All images include a light 'NeoGuard' watermark in the background. Figs 7.a and 7.b are two screenshots of the scoring page showing respectively the upper and lower views of a scored seizure which lasted 50 seconds.

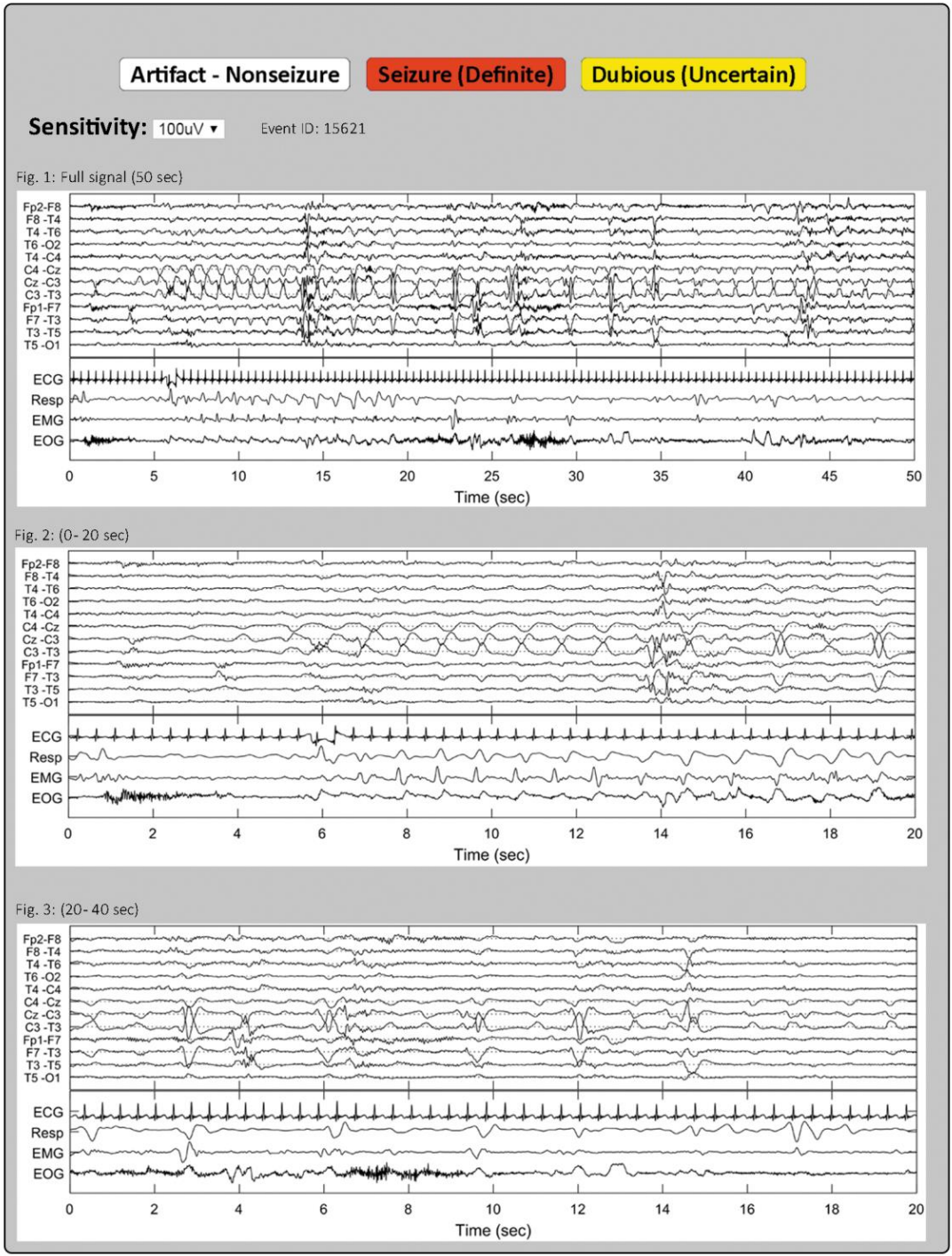

*Fig 7.a. The view of the upper part of the scoring page including the scoring buttons, and images 1-3. In each image, the EEG and polygraphic signals are shown.*

(16)

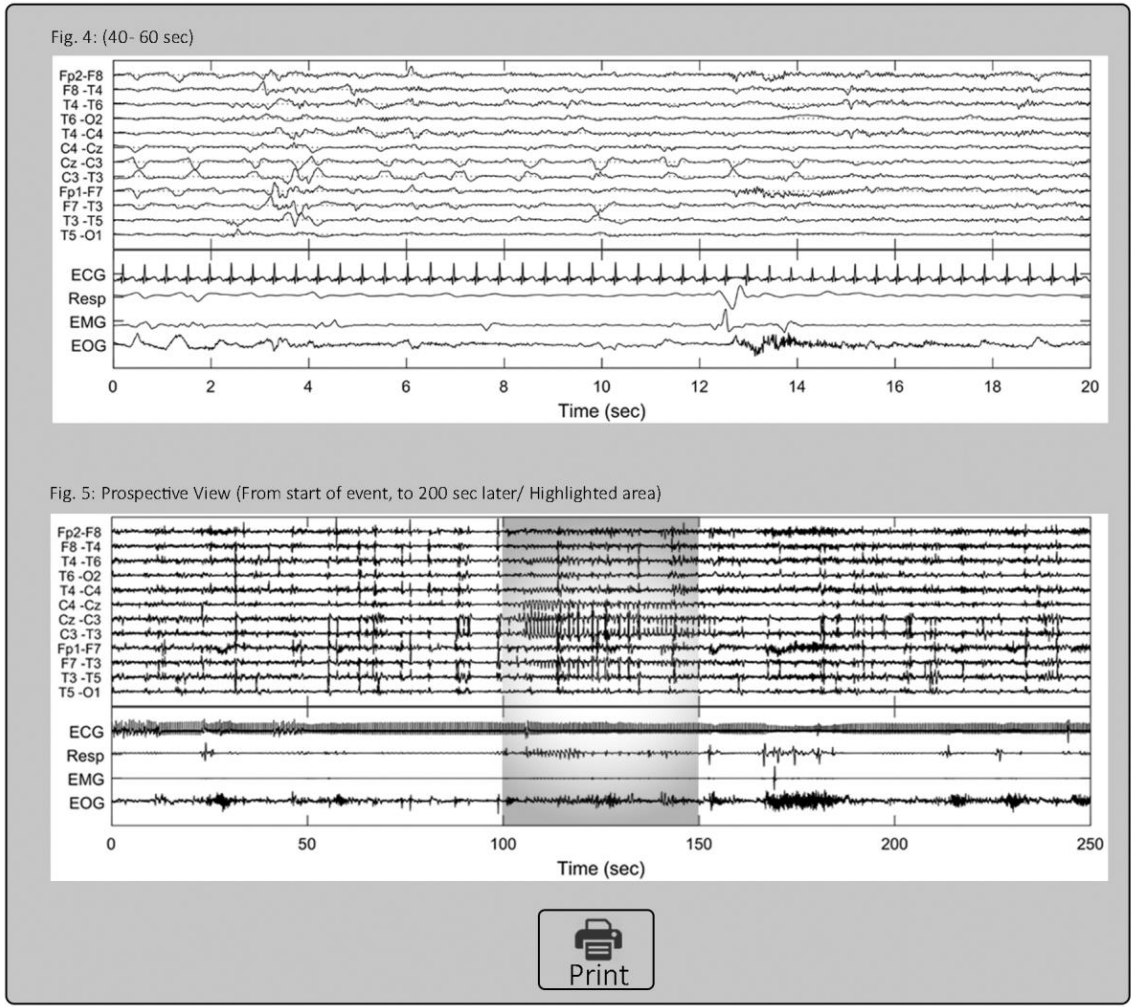

*Fig 7.b. The view of the lower part of the scoring page including the images 4-5 and the Print button. In each image the EEG and polygraphic signals are shown.*

# Results

When all events are scored by the user, the results page will be displayed with two tables. The first table shows all scorings in detail, with each row showing the user's choice of label and the corresponding scores of the experts. This allows the user to easily compare the scoring of each event. Furthermore, by clicking on the 'eye' icon in the first column of each row, one is able to review the images of the corresponding

(17)

event. The second table displays the agreement metrics, Joint-probability and Kappa, between the user and each of our expert readers. If the number of scored events is less than 60, a warning will be shown to indicate that the agreement metrics may not be accurate enough due to an insufficient number of scored events.

Fig 8 shows an example of a user who scored 10 events. In the first table, the green color shows the events with agreement between the expert readers and the user and the red ones shows events where they differ. The 'eye' icons which allow review of the corresponding EEG images, are located in the first column.

![Fig 8 screenshot]

Fig 8. A screenshot for the result of a user who scored 10 events

# Discussion

Several reports have shown that there is disagreement between expert EEG raters when they score neonatal seizures recorded on cEEG, particularly for lower amplitude patterns without very clear-cut evolution. This raises the issue that although there is a clear definition of neonatal seizure in theory, there is not yet a clear-cut gold standard for the interpretation of the EEG in practice, at least for the problematic presumed ictal patterns. Therefore, it seems essential for the EEG trainees who are learning to diagnose neonatal seizures to become familiar with the opinion of experts working in different centers. In this paper, a novel learning platform was introduced and explained. This tool can be used by EEG trainees to improve their abilities and by tutors in their classes, workshops, or seminars, to demonstrate practically how to recognize seizures.

One of the advantages of this online system is that the expert readers, and events can be updated or increased easily in the future. Although, the NGLP currently includes the scorings of four expert EEG readers from our participating centers, we are keen to include more EEG samples and volunteer expert EEG readers from other centers to enrich the database and the NGLP. The NGLP is able to connect to our database and automatically add newly recorded segments. Subsequently, as soon as the secondary readers score them, they will be available in the system for the NGLP users.

The other advantages of this system is that as soon as a statistically sufficient number of trainees have used this platform, extra analyses are possible on the trainees' scorings. For instance, the following questions can be addressed: what is the correlation between the level of experiment and agreement of the user? What is the correlation of wrong/true scorings with the duration of events, etiologies, agreement of experts, types of seizures, etc.?

However, the NGLP also has some limitations. Firstly, the users cannot watch the video of the neonates. As it is explained above, the gold standard of detecting seizures is visual inspection of cEEG-video. However, sharing the videos of neonates on the web is clearly not possible because of privacy and ethical issues. Secondly, the users are constrained by the provided EEG settings and cannot change the montage of the EEG signals or cannot move the window over time. These limitations may make the scoring process artificial and more difficult when compared to the normal scoring practice of cEEG used by experts. In addition, due to these limitations, the agreement measures could be underestimated. Nonetheless, the NGLP is the first platform which publicly shares the scorings of several expert EEG readers from different centers on the web. It is hoped that some of the limitations mentioned are overcome in the future versions of the NGLP.

Furthermore, in future versions, the users will have a personal profile for reviewing the previous scorings, sharing their knowledge and opinions with other users and experts, as well as customizing the website settings. Also, some interactive training tools and educational content can be added to the website.

## Intellectual property

The platform is publicly available at the NeoGuard website "www.neoguard.net". The users can freely copy, print, and distribute the images, events or the results of this platform, provided the author and original source are properly cited.